\documentclass[apj,numberedappendix]{emulateapj}
\usepackage{apjfonts}

\usepackage{graphicx}

\usepackage{amsmath}

\newcommand{\RL}{{R_L}} 
\newcommand{\RI}{{R_0}} 
\newcommand{\rg}{{r_g}} 
\newcommand{\psiout}{{\psi_L}}
\newcommand{\rhoout}{{\rho_{\text{out}}}}
\newcommand{\rhoin}{{\rho_{\text{in}}}}
\newcommand{\Mf}{{M_{\text{f}}}}
\newcommand{\vf}{{v_{\text{f}}}}
\newcommand{\nH}{{n_{\text{H}}}}
\newcommand{\vK}{{v_{\text{K}}}}
\newcommand{\vKo}{{v_{\text{Ko}}}}

\newcommand{\software}{\sc}

\newcommand{\B}{{\mathbf{B}}}
\newcommand{\ve}{{\mathbf{v}}}
\newcommand{\E}{{\mathcal E}}
\newcommand{\Ev}{{\boldsymbol{\E}}}
\newcommand{\jv}{{\mathbf{j}}}
\newcommand{\Efv}{{\mathbf{E}}}

\newcommand{\AU}{\,{\text{AU}}}
\newcommand{\cm}{\,{\text{cm}}}
\newcommand{\gram}{\,{\text{g}}}
\newcommand{\kelvin}{\,{\text{K}}}
\newcommand{\gauss}{\,{\text{G}}}
\newcommand{\kms}{\,{\text{km}\,{\text{s}}^{-1}}}
\newcommand{\yr}{\,{\text{yr}}}
\newcommand{\solarmass}{\,{\text{M}_\sun}}
\newcommand{\solarmassyr}{{\solarmass\yr^{-1}}}
\newcommand{\degree}{{\arcdeg}}

\newcommand{\citesq}[1]{\citeauthor{#1}\ [\citeyear{#1}]}

\begin{document}

\slugcomment{Submitted 2003 February 27; accepted 2003 June 13}

\journalinfo{The Astrophysical Journal, {\rm in press}}

\shorttitle{MAGNETO-CENTRIFUGAL LAUNCHING OF JETS: II}
\shortauthors{{KRASNOPOLSKY}, {LI} AND {BLANDFORD}}

\title{Magneto-Centrifugal Launching of Jets from Accretion
Disks. \break II. Inner Disk-Driven Winds}

\author{Ruben Krasnopolsky}
\affil{Center for Theoretical Astrophysics, University of Illinois at
Urbana-Champaign, Loomis Laboratory, 1110 West Green Street, Urbana, IL 61801}
\author{Zhi-Yun Li}
\affil{Astronomy Department, University of Virginia, Charlottesville,
VA 22903}
\and
\author{Roger D. Blandford}
\affil{California Institute of Technology, Mail Code 130-33,
1200 East California Boulevard, Pasadena, CA 91125}

\begin{abstract}
We follow numerically the time evolution of axisymmetric outflows driven
magneto-centrifugally from the inner portion of accretion disks, from
their launching surface to large, observable distances.
Special attention is paid to the collimation of part of the
outflow into a dense, narrow jet around the rotation axis,
after a steady state has been reached. For parameters typical
of T Tauri stars, we define a fiducial ``jet'' as outlined by
the contour of constant density at $10^4\cm^{-3}$. We find
that the jet, so defined, appears nearly cylindrical well
above the disk, in agreement with previous asymptotic analyses.
Closer to the equatorial plane, the density contour can either
bulge outwards or pinch inwards, depending on the conditions at
the launching surface, particularly the mass flux distribution.
We find that even though a dense, jet-like feature is always
formed around the axis, there is no guarantee that the high-density
axial jet would dominate the more tenuous, wide-angle part of the
wind. Specifically, on the $100\AU$ scale, resolvable by HST
and ground-based adaptive optics for nearby T Tauri winds,
the fraction of the wind mass flux enclosed by the fiducial
jet can vary substantially, again depending on the launching
conditions. We show two examples in which the fraction is
$\sim 20\%$ and $\sim 45\%$. These dependences may provide a way to
constrain the conditions at the launching surface, which
are poorly known at present.
\end{abstract}
\keywords{galaxies: active --- ISM: jets and outflows ---
methods: numerical -- MHD --- stars: formation}

\section{Introduction}

\subsection{The Quest for Large-Scale Wind Structure}

In a previous paper
(\citealt{Krasnopolsky99}; hereafter Paper~I),
we described a modification of the {\software{Zeus}} MHD code
and its application to
the problem of launching a magneto-centrifugal wind from a Keplerian
disk. The disk was
idealized as a boundary condition at the equatorial plane,
on which open magnetic fields are firmly anchored {\it at all radii}
inside the simulation box. Mass was injected supersonically but
sub-Alfv{\'e}nically onto the open field lines at a prescribed rate
and accelerated magneto-centrifugally along the field lines to produce
a high speed wind. Our approach followed that of
\citet{Ustyugova95} and \citet{Ouyed97}.
It permitted a decoupling of the dynamics
of the wind from that of the disk and allowed for a thorough investigation
of the structure of the wind. Previous numerical studies of
magneto-centrifugal winds were limited to a relatively small region
not far from the launching surface. In this paper, we seek to extend
the wind solutions to large, observable distances, and to relate the
observable flow quantities on large scales to the launching conditions.

\begin{figure}[tb]
\plotone{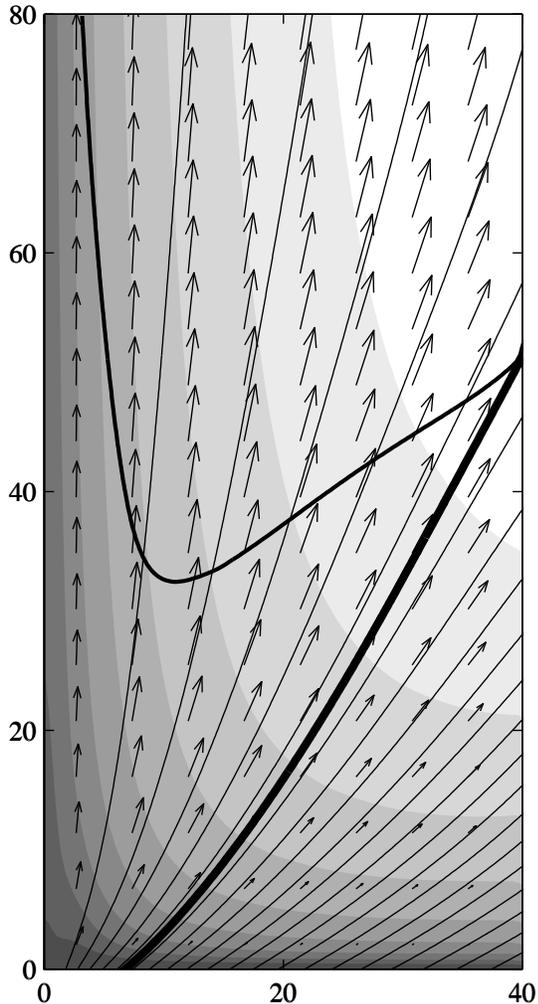}
\caption{A representative magneto-centrifugal wind launched from a
Keplerian disk (in arbitrary units; taken from Paper~I). Shown are
magnetic
field lines (light solid lines), velocity vectors (arrows) and the
fast magnetosonic surface (solid line of medium thickness). The
thickest solid line
divides the portion of the wind that becomes super fast-magnetosonic
inside the simulation box (above) from the portion that does not
(below).}
\label{fig:1}
\end{figure}

To study the large-scale structure of a disk-driven magneto-centrifugal
wind numerically, a technical problem must first be overcome. This
problem is illustrated in Fig.~\ref{fig:1} (taken from Paper~I),
where a  disk-wind solution representative of those found in existing
time-dependent simulations (e.g., \citealt{Romanova97};
\citealt{Ouyed97};
\citealt{Ustyugova99}) is shown. Note that a large fraction of the
wind remains sub--fast-magnetosonic in the computational realm.
Information from the edge of the simulation box can propagate
upstream in a sub--fast-magnetosonic region, creating an undesirable
coupling between the box boundaries and the wind-launching.
This coupling, we believe, is at the root of the sensitive dependence
of wind solutions on the size and shape of the simulation box
found previously by us and others (see \S\ref{implementation}
and \citealt{Ustyugova99} for a discussion).
In some cases, merely changing the shape of the box
can destabilize a wind and induce chaotic flow fragmentation. This
dependence must be eliminated before one can study the large scale
structure of the magneto-centrifugal wind with confidence.

One way to overcome the above difficulty is to restrict the wind
launching to the inner region of an accretion disk. In the specific
context of star formation, there are reasons
to believe that the bulk of the outflow is indeed driven from an
inner disk. There are two popular magneto-centrifugal models for the
production of optical jets, such as the spectacular HH 30 jet
studied in detail with the HST (\citealt{Burrows96}; \citealt{Ray96}).
The models differ on
where the wind-driving open magnetic fields are anchored. If the
open field lines are anchored in a narrow region of disk near
the corotation radius of the stellar magnetosphere, then the wind is
called an ``X-wind'' \citep{Shu00}. The X-wind serves the fundamental
purpose of removing angular momentum from the central star and keeping
it rotating at a rate well below the breakup rate, as observed. If,
on the other hand, the field lines are anchored over a wider region
of the disk, then a ``disk-wind'' is produced (eg., \citealt{Konigl00}).
The disk-wind may be primarily responsible for driving the
mass accretion through the portion of the disk where the wind is
launched.
Wind launching should be easier from the inner part of
protostellar disks, where the temperature is high enough ($T> 10^3\kelvin$)
that thermal ionization of alkali metals ensures that the disk material
and wind-launching magnetic field are well coupled. In the cooler outer
part, the bulk of disk material is too weakly ionized to participate
in the wind launching, although in the surface layer and at large radii
(say $\sim 10^2\AU$; \citealt{Wardle93}) protostellar X-rays and
Galactic cosmic rays (if not
excluded by the wind) may provide a level of ionization sufficient
for good coupling.
We will not consider in this paper the possible outer disk wind, which
is probably less relevant for the fast jets and winds of hundreds of
$\kms$ that we are interested in. Such a wind can modeled with our
simulation setup (see \S\ref{implementation}) if desired since the size
of the launching region is fully adjustable.

The large-scale structure of magneto-centrifugal winds is a prerequisite
for quantitatively modeling a variety of outflow phenomena observed
around YSOs, including optical jets, neutral atomic winds, and bipolar
molecular outflows. It is the observational data on these phenonema
that will constrain the conditions at the wind launching surface, which
are poorly known. Particularly tight constraints are
expected from high spectral as well as high spatial resolution observations
of forbidden line emission from T Tauri winds. A recent
example of such observations is that of \citet{Bacciotti00}. They
constructed synthetic two-dimensional images of the DG Tau wind based
on HST/STIS data in several forbidden lines (including [OI]$\lambda
\lambda$6363,6300, [SII]$\lambda$6731, and [NII]$\lambda$6583) and
H$_\alpha$ in several radial velocity intervals, with a spatial
resolution of order $0.1\arcsec$ (or about $15\AU$ at the
distance of DG Tau). To take full advantage of such observations, one
needs a reliable model that can follow the flow from the (small)
launching surface to observable distances. As a first step towards
such a model, we restrict our attention to those magneto-centrifugal
winds that can reach a steady state, even though most observed jets
are time-dependent, possibly episodic \citep{Reipurth99}. We
postpone an investigation of the intrinsically time-dependent winds
to a future paper.

Although our intended application is primarily to YSOs, we note that
magneto-centrifugal winds driven from a limited region of accretion
disks may occur in other astrophysical systems, e.g., cataclysmic
variables (CVs) and possibly planetary nebulae. A related jet launching
mechanism, through torsional Alfv{\'e}n wave, was first discussed in
\citet{Uchida85}, and refined in many subsequent works (see,
e.g., \citealt{Kato02} and references therein).

\subsection{Uncertain Jet Formation Efficiency}

Asymptotic structure of steady magneto-centrifugal winds has been analyzed
by \citet{Heyvaerts89} and \citet{Shu95}, among others. The
focus of such analyses was on jet formation, through the hoop-stress
associated with toroidal magnetic field. While there is a general agreement
that a dense, narrow jet can indeed form in the axial region of a
magneto-centrifugal wind (see, however, \citesq{Okamoto99}
for a different view),
it is not clear how the jet properties, such as their density, mass and
momentum fluxes, depend on the launching conditions. Indeed, the term
``jet'' was used by different authors to mean different things.
\citet{Bogovalov99}, for example, refer to the axial region with a
more or less uniform density and magnetic field strength and vertical
streamlines as the jet, whereas \citet{Shu95} suggested that the
observed jet is simply the dense part of an underlying anisotropic wind
above, roughly speaking, some fiducial density that shows up more prominently
than the more tenuous equatorial part in the emission of forbidden lines
such as [OI] and [SII] \citep*{Shang98}.
We adopt the latter meaning
of ``jet'', which is more observationally oriented. For simplicity, the
fiducial density at the jet boundary is chosen to be $\nH=10^4
\cm^{-3}$, comparable to the values derived by
\citet*{Bacciotti99}
for the famous HH 30 jet\footnote{We note that the density derived
by \citet{Bacciotti99} for the HH 30 jet is not constant along the
jet, but decreases from a value $\sim 10^5\cm^{-3}$ close to the exciting
source to a value $\sim 10^4\cm^{-3}$ further out.}. An important
question is: what fraction of the mass flux of the wind is collimated
into the fiducial jet?
We will address this question using numerically
determined, large-scale wind solutions.

The fraction of mass flux residing in the jet measures, in some sense, the
efficiency of jet formation. It can be constrained by observations. Take
the HH 30 system as an example. For this object, one can eliminate a jet
formation efficiency as low as $10\%$ from the following considerations.
\citet{Bacciotti99} showed that the mass flux in the main
jet is $\sim 1.7\times 10^{-9}\solarmassyr$ ($\sim 1.8\times 10^{-9}
\solarmassyr$ in the counter jet)
close to the exciting source. If only $10\%$ of the mass
flux resides in the jet, the total mass flux of the wind from both sides
would be $\sim 3.5\times 10^{-8}\solarmassyr$. If a third
\citep{Shu95} or less
(e.g., \citealt{Ferreira95}) of the material accreted
through the disk gets ejected in the wind, then the disk mass accretion
rate must be $\sim 10^{-7}\solarmassyr$ or higher. Such a large
accretion rate would be difficult to accommodate by the very weak IR
excess from ISO observations \citep{Stapelfeldt98}, which points
to a relatively inactive disk, and by the low disk mass
($\sim 10^{-3} \solarmass$) estimated from scattered light and dust
emission \citep{Stapelfeldt01},
although this estimate is uncertain. Increasing
the jet mass flux fraction to $50\%$ would lower the disk accretion rate
to $\sim 2\times 10^{-8}\solarmassyr$ (a more typical value for T Tauri
disks; \citealt{Gullbring98}), if a third of the accreted material is
ejected. If a smaller fraction of the accreted material is ejected, then
the required accretion rate would be higher. It therefore appears that, at
least for HH 30, the jet formation efficiency must be fairly high, of
order $50\%$ or more. It is not clear whether such a high efficiency
can be achieved naturally in a magneto-centrifugal wind.

The rest of the paper is organized as follows. In \S\ref{formulation}, we
present our formulation of the problem of launching magneto-centrifugal
winds from only inner disk regions. It is followed by a discussion of
a ``standard'' wind solution which illustrates several generic features
of jet formation and large-scale structure of steady magneto-centrifugal
winds in \S\ref{standard}. In \S\ref{load}, we examine the effects of
the mass flux distribution at the launching surface on the jet/wind
properties. We conclude and discuss future work in \S\ref{conclusion}.

\section{Formulation of the Inner Disk-Driven Wind Problem}
\label{formulation}

\subsection{Implementation of a Finite Launching Surface}
\label{implementation}

\begin{figure}[tb]
\plotone{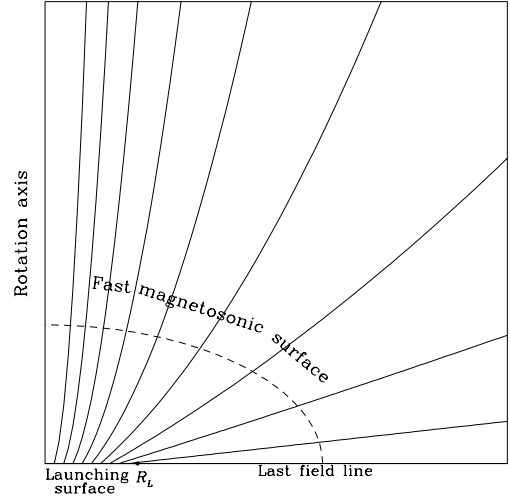}
\caption{Schematic view of a magneto-centrifugal wind launched from a
limited, inner disk region.}
\label{fig:2}
\end{figure}

As mentioned earlier, we limit the launching of magneto-centrifugal
winds from the inner, relatively hot regions of protostellar disks.
Numerically, this restriction
enables us to extend the wind solution to large distances
without having to worry about the undesirable coupling between the
simulation box and the launching region, which occurs in the sub
fast-magnetosonic region in the lower-right corner of the
solution shown in Fig.~\ref{fig:1}. The reason for that region of flow
to remain sub fast-magnetosonic is simple: the wind coming off
the outer part of the disk encounters the edge of the simulation
box too soon. In other words, the wind in the region simply
does not have enough room to get accelerated to the fast magnetosonic
speed. This situation remains true as long as the wind is driven
off {\it all} of the (equatorial) disk plane (as assumed in all
previous time-dependent disk-wind simulations), regardless of the
box size. This troublesome feature can be eliminated by restricting
the wind launching to only the inner region of a disk (within a
radius denoted by $\RL$), as sketched in Fig.~\ref{fig:2}. To fill
all available
space above (and below) the equatorial plane, we demand that the
last field line lie exactly on
the equatorial plane\footnote{Of course, if the wind is confined
externally, either by a flaring disk or an ambient medium, the last
field line must be modified accordingly. External
confinement of magneto-centrifugal winds is an important topic
that we plan to explore numerically in the future. It is
expected to play a more dominant role in the outflows of embedded
sources than in the winds of optically revealed T Tauri stars.
Note that the dense material inside the disk beyond the wind launching
region prevents the ``last'' field lines of both sides of the disk from
reconnecting. Also, the interface between the wind and disk
material outside $\RL$
may become Kelvin-Helmholtz unstable. Such additional
complications are ignored in this paper.}.
Wind plasma sliding along the last (horizontal) field line,
anchored at the radius $\RL$, will become super fast-magnetosonic in
the computational realm, provided that the size of the simulation box
is sufficiently large. Once the whole fast magnetosonic
surface is completely enclosed inside the simulation box, the size
and shape of the box should have
minimal effects on the structure of the wind, especially near
the launching surface, since information cannot propagate upstream
in a super fast-magnetosonic region. In this way, we should be
able to study the wind structure up to arbitrarily large distances from
the source region, limited only by computer time.

There is, however, an additional consideration. \citet{Ustyugova99}
have shown that artificial dependence of simulation results on
computational box shape can appear even in a
super--fast-magnetosonic flow, if the Mach cones at the
boundaries intersect the simulation region. The
opening angle $\varphi$ of the Mach cone around a streamline
is given (for a cold flow) by $\tan^2\varphi=1/(M_f^2-1)$,
where $M_f$ is the fast Mach number.  As the flow becomes more
super--fast-magnetosonic, the cone gets narrower, more
forward-directed, and has less
chance of intersecting the simulation region. Our
simulations, designed to produce clearly super-fast flows,
fulfill this requirement everywhere except for a small region
around the axis. As we discuss below, the axial region
receives a special treatment, and does not appear to create
any numerical problems.

\subsection{Equations}

For completeness and to define symbols, we list below the time-dependent
MHD wind equations that we solve numerically:
\begin{eqnarray}
\frac{\partial\rho}{\partial t}+\nabla\cdot\left(\rho\ve\right)&=&0\\
\rho\frac{\partial\ve}{\partial t}+\rho\left(\ve\cdot\nabla\right)\ve
&=&-\nabla p+\rho\nabla\Phi_g+\jv\times\B/c\\
\label{dBdt}
\frac{\partial\B}{\partial t}&=&\nabla\times\left(\ve\times\B\right)=
\nabla\times\Ev\\
\frac{\partial u}{\partial t}+\nabla\cdot\left(u\ve\right)
&=&-p\nabla\cdot\ve\\
p&=&(\gamma-1)u
\end{eqnarray}
where
\begin{eqnarray*}
\rho&=&\text{matter density}\\
\ve&=&\text{velocity flow field}\\
\B&=&\text{magnetic field}\\
\jv&=&(c/4\pi)\nabla\times\B=\text{current density}\\
\Ev&\equiv&\ve\times\B=-c\Efv\ \  \text{where}\ \  \Efv
\ \ \text{is the electric field} \\
\Phi_g&=&\text{gravitational potential}\\
p&=&\text{thermal pressure}\\
u&=&\text{internal energy density (per unit volume)}\\
\gamma&=&\text{adiabatic index}
\end{eqnarray*}
We have adopted a cylindrical coordinate system $(R,\phi,z)$, with
the spherical radius given by $r=\sqrt{R^2+z^2}$.
The gravity field is smoothed inside a sphere of radius $\rg$,
through
\begin{equation}
\Phi_g=\frac{\Omega_0^2}{2} (r^2-3 \rg^2),
\label{smoothing}
\end{equation}
where $\Omega_0$ and $\rg$ are fixed parameters, whose product $\vKo
=\Omega_0 \rg$ gives the scale for the rotational speed
$\vK\equiv\sqrt{R\left(\partial\Phi_g/\partial R\right)_{z=0}}=
\Omega_0 R$ on the disk inside $\rg$. Outside $\rg$, the gravity
is not smoothed, and the potential has the usual form
\begin{equation}
\Phi_g=-\frac{\Omega_0^2 \rg^3}{r}\ .
\label{grav}
\end{equation}
It yields a Keplerian speed of
\begin{equation}
\vK=\vKo\left(\frac{\rg}{R}\right)^{1/2}
\label{kep}
\end{equation}
on the disk.
The parameter $\Omega_0$ is simply the angular speed at the inner edge
of the (unsoftened) Keplerian disk. We label magnetic field lines
with the magnetic flux function
$\psi$ defined by $\partial\psi/\partial R=RB_z$. Except
for a few specific differences noted here and below, the notations
and numerical methods we use are the same as those in Paper~I, to
which we refer for more details.

The governing equations are solved numerically using
the {\software{Zeus}} code, subject
to a set of appropriate initial and boundary conditions. We use a
parallelized version of the {\software{Zeus3D}} code (\citealt{Clarke94}),
whose main algorithms are based on \citet{Stone92a,Stone92b} and
\citet{Hawley95}.
The magnetic field is evolved by the well-known constrained
transport method \citep{Evans88},
which preserves the condition $\nabla\cdot\B=0$
at all times as long as this condition is satisfied everywhere
initially.

\subsection{Boundary Conditions}
\label{bc}

On the four sides of our 2D simulation box (see Fig.~\ref{fig:2}), we need
to impose boundary conditions. Numerically, these conditions are imposed
by assigning values to flow quantities in a few ghost zones just outside
the active computational grid. The ghost zone quantities include the
density $\rho$, internal energy $u$, the three components of velocity
$\ve$, and the three components of the field $\Ev\equiv\ve\times\B$.

The conditions on the axis $R=0$ and the outer boundaries $R=R_{\max}$
and $z=z_{\max}$ are relatively straightforward: on the axis we
demand reflection symmetry, whereas on the outer boundaries we allow
for outflow. These conditions are standard options for the {\software{Zeus}}
code. The conditions on the equatorial plane
$z=0$ are more difficult to specify and implement. They represent
the main technical improvement that distinguishes this work from others.

For our problem, the $z=0$ surface is divided into two distinct regions
on which different boundary conditions are imposed (see Fig.~\ref{fig:2}).
Inside the radius $\RL$, which divides these two regions, magnetic field
lines are anchored and outflowing materials are injected. The boundary
conditions there are the same as those used in Paper~I. Specifically,
the quantities $\rho$, $B_z$ and $v_z$
are specified as known functions of the radius $R$ at all times,
and the two horizontal components of the $\Ev$ field satisfy $\E_\phi=0$
and $\E_R=\vK B_z$. To implement these boundary conditions for
$\Ev$, the horizontal components of $\ve$ in the ghost zones
are chosen so that
$v_R/v_z=B_R/B_z$, and $v_\phi=\vK+B_\phi v_p/B_p=\vK+B_\phi v_z/B_z$.
In the ghost zones below $z=0$, we force the components of $\Ev$ to
have values: $\E_\phi(-z)=-\E_\phi(z)$,
$\E_R(-z)=2\vK(R) B_z(R)-\E_\phi(z)$, and $\E_z(-z)=\E_z(z)$.
These conditions allow the foot point of a wind-launching open field
line to be pinned firmly on the
launching surface and at the same time leave the field line free
to bend both radially and azimuthally in response to the stresses
exerted by the wind material.

Our boundary conditions on the launching surface inside $\RL$ are similar
to those used by \citet{Ustyugova95,Ustyugova99} and
\citet{Romanova97}, except that they also
considered the case of subsonic wind injection. In the subsonic case, the
wind must pass smoothly through a sonic point close to the launching
surface, which fixes the mass loading rate and reduces the number of the
quantities that must be specified at the $z=0$ surface by one. The sonic
point is sensitive to the thermal and magnetic structure of the disk
atmosphere, which is not well understood. We will postpone a detailed
treatment of the sonic transition to a future investigation.
\citet{Ouyed97} appear to have fixed all flow quantities in
the ghost zones, which is different from our treatment. Fixing all
quantities in the ghost zones risks creating discontinuities in the
radial and toroidal components of the magnetic field between the ghost
and active zones, since in the active zones these two components cannot
be prescribed; they are determined by the stresses in the wind.

Outside the radius $\RL$, the equatorial $z=0$ surface is occupied
by the last flux surface along which the wind material emanating from
the radius $\RL$ flows. We demand that the vertical field
component $B_z=0$ on this portion of the $z=0$ surface, and
$B_z(-z)=-B_z(z)$ in the ghost zones below $z=0$. These conditions
on the magnetic field are enforced by the conditions on the field $\Ev$,
which are $\E_\phi(-z)=-\E_\phi(z)$, $\E_R(-z)=-\E_R(z)$, and
$\E_z(-z)=\E_z(z)$.  Equation (\ref{dBdt}) then ensures that the initial
value of $B_z$ at $z=0$ will not change.
For the hydrodynamic quantities, we demand that
$\rho$, $u$, $v_R$ and $v_\phi$ be continuous across the $z=0$
surface and assign the value $v_z=0$ to the $z=0$ surface and the
ghost zones below.

To complete our discussion of boundary conditions on the $z=0$ surface,
we note the following three additional features. First, to fill out
space, the field lines anchored in the wind launching region must
become more and more horizontal as the edge of the region (at $R=\RL$)
is approached. This condition is satisfied by forcing the functions
defining $v_z(R)$ and $B_z(R)$ for $(z=0,\,R<\RL)$ to zero at $R=\RL$,
keeping their ratio finite. Second, to prevent inflow in the region
surrounding the rotation axis where the magneto-centrifugal mechanism
fails, we inject a
cold, light, fast axial flow capable of escaping
the gravitational pull of the central compact object. Plausible
physical justifications for this fast injection are given in Paper~I.
Finally, during the wind acceleration, there can appear temporary
negative values of $v_z$ in active zones next to the launching disk.
For this case only,
we allow the boundary condition at $z=0$ to absorb the flow,
by setting $v_z$ to zero at $z=0$, and allowing the other hydrodynamic
quantities to have the same value in the ghost and active zones,
but not altering the conditions on the $\Ev$ field.  This
procedure avoids some numerical defects, as explained in Paper~I.
The need for this special treatment disappears once a steady state near
the disk surface is reached.

\subsection{Initial Magnetic Field Distribution}
\label{potential}

Following \citet{Ouyed97}, we choose as the initial (purely
poloidal) magnetic field distribution above the disk a current-free
field compatible with the function $B_z(R)$ defined at $z=0$. Such
an initial field has the advantage of perturbing the ambient
medium as little as possible. (This gas will eventually be pushed outside the
computational domain by the fast moving wind launched from the disk.)
One can solve the force free equation $\nabla^2\psi=0$
for the flux function
$\psi$ through separation of variables, with the following set of
boundary conditions: $\psi(R=0,z)=0$, $\psi(R,z=\infty)=0$,
$\psi(R=\infty,z)=0$, and
$\psi(R,z=0)=\psi_D(R)$. The function $\psi_D(R)$ is prescribed
in the launching region inside the radius $\RL$; beyond $\RL$, it
has a constant value $\psiout$. A formal solution that satisfies
these conditions is
\begin{eqnarray}
\psi(R,z) &=& R \int_0^\infty  dk\, e^{-kz} J_1(kR)\, g(k)\\
g(k)  &=&    k \int_0^\infty  dR\, J_1(kR)\, \psi_D(R)\\
      &=&    k \int_0^{\RL} dR\, J_1(kR)\, \psi_D(R)
           +                     J_0(k\RL)\, \psiout
\end{eqnarray}
where $J_0$ and $J_1$ are Bessel function of zeroth and first order.
Doing these integrations at each grid
point can be expensive, especially in regions where the integrals
are slow to converge, due to the rapidly oscillatory integrand.
A more convenient and surer procedure to find the solution consists
of first evaluating
with sufficient approximation the integrals at the outer boundaries
$R=R_{\max}$ and $z=z_{\max}$, and then using these values to solve an
elliptic problem with Dirichlet boundary conditions, together with the
known values of $\psi_D$ at the disk and axis. Since the size
of the simulation box is much larger than the size of the
launching region, one can expand the integrals in powers of $k$
at distances far from the launching region (i.e., $r\gg \RL$).
The first two terms of expansion, which proved sufficient to
calculate our boundary values, are
\begin{equation}
\psi(R,z) = \psiout \left(1 - \frac{z}{r}\right) +
\frac{3 C z R^2}{r^5},
\end{equation}
where $C = \int_0^{\RL} R\,\psi_D(R)\,dR - \psiout \RL^2/2$.
We solved the elliptic equation by a relaxation method,
taking as an initial guess
the simple distribution $\psi(R,z) = \psi_D(r)(1 - z/r)$.

With the governing equations, numerical method, initial and boundary
conditions explained, we are ready to examine in detail a
representative wind solution that has reached a steady state.

\section{An Illustrative Wind Solution}
\label{standard}

\subsection{Simulation Setup}

We first establish a ``standard'' run against which other simulations
will be compared. All simulations are carried out with dimensionless
quantities, but to facilitate comparison with high resolution
observations that are becoming available, especially from the HST,
we will present results with dimensional units. To fix units, we
adopt one solar mass for the central object $M_\ast$, and $\RI=0.1\AU$
for the inner radius of the wind-launching region,
which is roughly the corotation
radius of the stellar magnetosphere (inside which the disk is either
very sub-Keplerian or non-existent; \citealt{Konigl91}; \citealt{Shu95}).
The gravitational softening radius $\rg$ is set equal to $\RI$.
The central mass and inner radius set a scale for the velocity,
which is the Keplerian speed at $\RI$, $\vKo=\sqrt{GM_\ast/\RI}
=94\kms$. The scales for the magnetic field strength,
$B_0$, and the density, $\rho_0$, are related by the requirement
that $B_0^2/(4\pi\rho_0)=\vKo^2$. Both of these scales are
fixed by demanding that the total mass flux of the outflow ${\dot M}$
(from each side of the disk) be ${\dot M}_0=10^{-8}\solarmassyr$, a value
representative for classical T Tauri stars (e.g., \citealt{Edwards93}).

For the standard run, we choose an outer radius for the wind launching
region, $\RL=1\AU$, ten times the inner radius $\RI$.  On the
equatorial plane $z=0$ inside $\RL$, one needs to
specify the distributions of the density $\rho$ at the base of the
wind, and the vertical components of both the injection velocity ($v_z$)
and the magnetic field strength ($B_z$). We consider a single-component
distribution for the magnetic field
\begin{equation}
B_z(R)={B_{z,\max}}{\left[1+\left({R/\RI}\right)^2\right]^{-e_b/2}}
\times f(R),
\end{equation}
where the parameter $B_{z,\max}$ sets the scale for the field strength and
the exponent $e_b$ the distribution. They are taken to be
$4\,B_0$ and $1.5$ respectively for the standard run (where $B_0$ is the
magnetic field unit defined above). The function $f(R)$ is set to
unity between the origin and a transitional radius $R_T$, and
\begin{equation}
f(R)=\left(1+2\frac{R^2-R_T^2}{\RL^2-R_T^2}\right)
      \left(\frac{\RL^2-R^2}{\RL^2-R_T^2}\right)^2
\label{spline}
\end{equation}
between $R_T$ and the outer radius of the launching region $\RL$. It is
used to force both $B_z$ and $v_z$ to vanish together at the outer edge
of the launching region, as required by the boundary conditions (see
\S\ref{bc}). We choose $R_T=0.8\AU$.

For the injection speed at the launching surface, we consider a two-component
distribution, with
\begin{equation}
v_z(R)=\frac{v_{z0}}{1+\left({R/\RI}\right)^2}
\end{equation}
in the fast injection axial region inside $\RI$, and
\begin{equation}
v_z(R)=\alpha\ \vK\times f(R)
\end{equation}
in the magneto-centrifugal wind launching region between $\RI$
and $\RL$. The
parameter $v_{z0}$ sets the maximum injection speed on the axis, and is
taken to be $160\kms$ for the standard run. The parameter $\alpha$
denotes the ratio of the injection speed to the Keplerian speed,
and is set to a small value of 0.1 for the standard run.

For the density distribution at the base of the wind $z=0$, we adopt
a small, constant value of $\rhoin=0.1\,\rho_0$ (where $\rho_0$
is the density unit) in the fast injection region inside the radius
$\RI$. On the Keplerian disk between $\RI$ and $\RL$, we prescribe a
power-law distribution
\begin{equation}
\rho(R)=\rhoout\left(\frac{\RI}{R}\right)^{e_\rho},
\label{deninj}
\end{equation}
up to the outer edge of the launching region $\RL$. The parameters $\rhoout$
and $e_\rho$ set the density scale and its decline rate with
radius. They are taken to be $\rhoout=10\,\rho_0$ and $e_\rho=1$
for the standard run. Note the factor of $10^2$ contrast between the
densities of the fast axial injection ($\rhoin$) and the inner
edge of the magneto-centrifugal wind ($\rhoout$).

\begin{figure*}[tb]
\plottwo{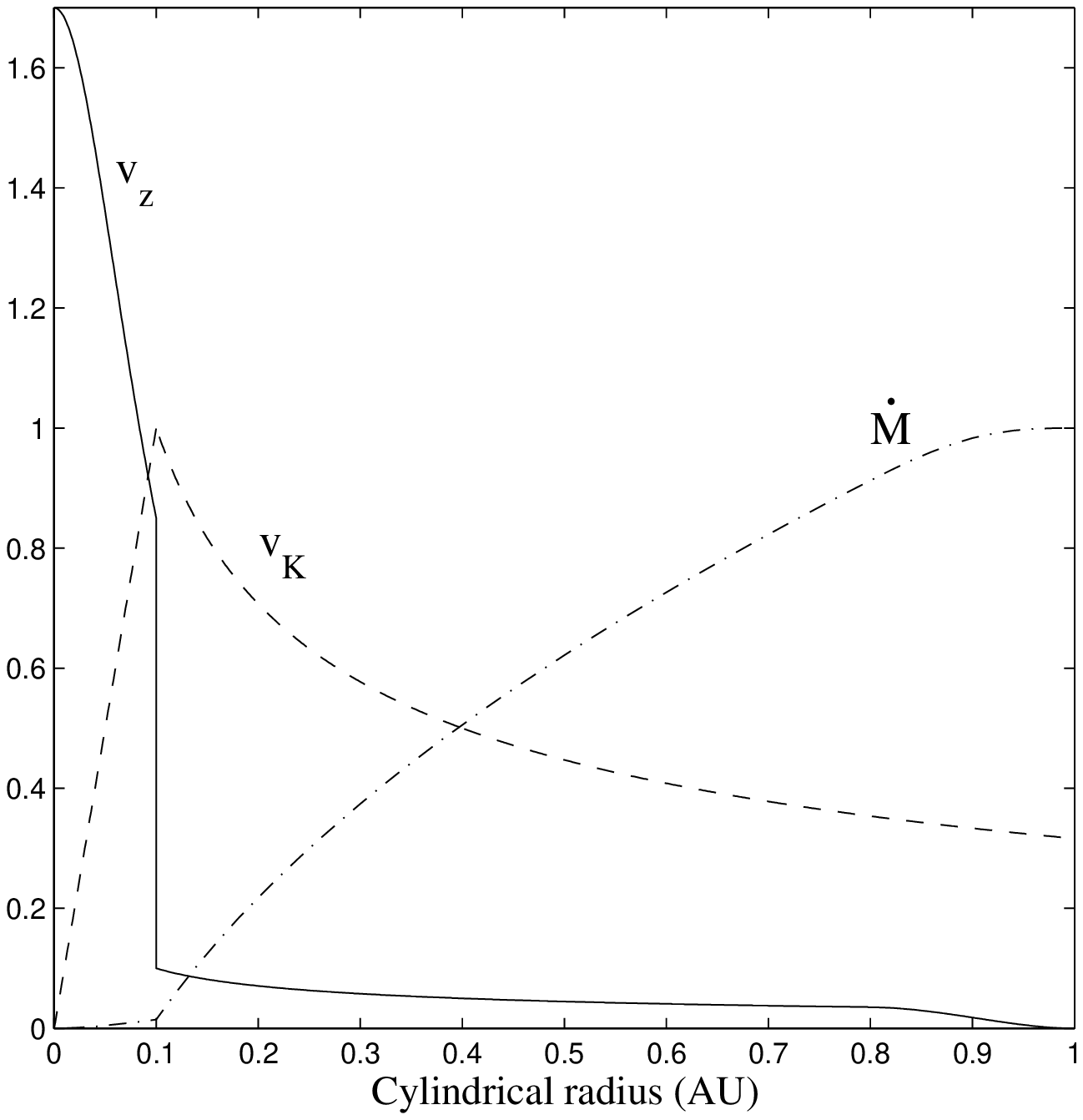}{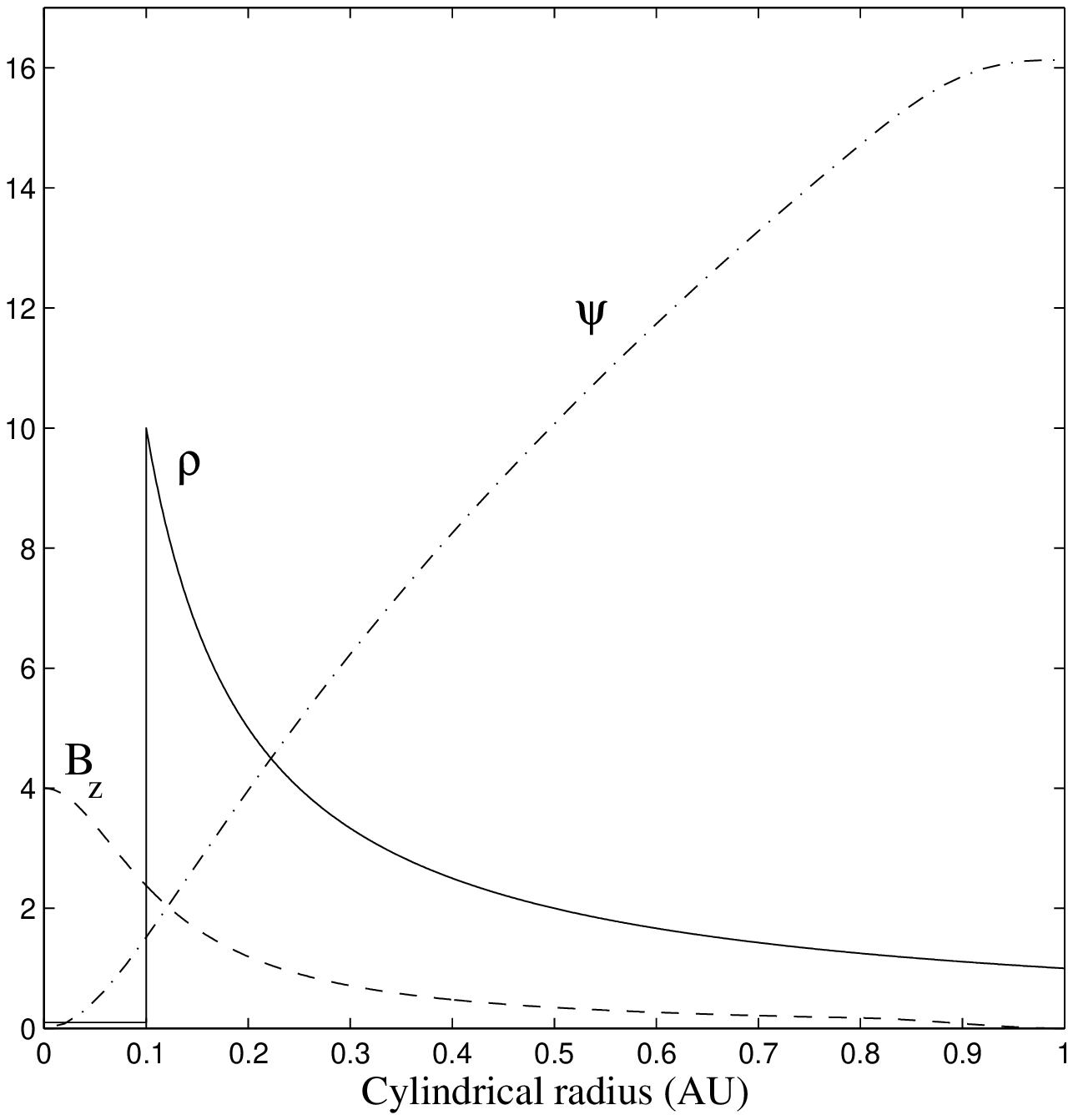}
\caption{The conditions specified on the wind launching surface $z=0$ for
the standard run in dimensionless units. The dimensional units are
$\vKo=94\kms$ for velocity, ${\dot M}_0= 10^{-8}\solarmassyr$ for mass
flux, $B_0=1.1\gauss$ for field strength, and $\rho_0=1.2\times 10^{-15}
\gram\cm^{-3}$ for density. The magnetic flux $\psi$ has a unit of
$B_0 \RI^2$, where $\RI=0.1\AU$ is the length unit.}
\label{fig:3}
\end{figure*}

The launching conditions specified above yield the following units for
the field strength and density: $B_0=1.1\gauss$ and $\rho_0=1.2\times
10^{-15}\gram\cm^{-3}$. These conditions are displayed in Fig.~\ref{fig:3},
together with a distribution of the accumulative mass flux injected
into the wind and the magnetic flux enclosed within a given radius on the
disk. It is clear that the fraction of mass flux in the fast injection
region inside $R=\RI$ is small ($\sim 1.5\%$), despite the large injection
speed in this region; the large speed is more than offset by the low
density specified. The large velocity shear near $R=\RI$ does not appear
to create any instability, possibly because of the strong (poloidal) magnetic
field present in the region.

Above the equatorial surface $z=0$, we start with a purely poloidal,
potential magnetic field, constructed using the method outlined
in \S\ref{potential}. The computational domain is initially filled
with a low density medium, whose density distribution is
prescribed either analytically or taken from other, similar runs
that have reached a steady state. Judicious choice of the initial
ambient density distribution can speed up the flow convergence to
the steady state enormously, although the final wind solution does
not depend on the choice, since the ambient medium will eventually
be swept out of the simulation box completely. To isolate the
magneto-centrifugal wind
acceleration and collimation, the focus of our study, from thermal
effects, we keep the flow cold at all times by arbitrarily setting
a small value of $0.02\kms$ for the sound speed.

Our main goal is to study the structure of magneto-centrifugal winds
on large, observable scales. For this purpose, we adopt a simulation
box of $100\AU \times 100\AU$. It is $100$ times the size of
the wind launching region and $1000$ times the inner Keplerian disk
radius $\RI$. Such a scale is directly accessible to high resolution
observations using the HST and ground-based adaptive optics
\citep{Dougados00} for nearby optically revealed, T Tauri winds.
Outflows from deeply embedded sources can be probed on a similar scale using
water masers. For example, the water masers of the Class 0 source S106
FIR appear to lie on a $3\AU$ (width) $\times$ $4\AU$ (length) U-shaped
surface some $25\AU$ away from the central source, possibly driven by a
well-collimated ``micro jet'' \citealt{Furuya99}).  Numerically, we
cover the active computation domain with $190{\times}210$ zones.
In the vertical direction, there are 40 uniform grid points covering
the region from 0 to $0.8\AU$, and 150 non-uniform zones covering
the region from 0.8 to $100\AU$, with the zone size increasing by a
small factor of 1.035023 between adjacent zones when moving outwards.
In the horizontal direction, the first 60 grid points from 0 to $1.2\AU$
(thus covering the entire launching region)
are uniform, followed by 150 non-uniform grid points with
approximately the same ratio between zones as before (1.034987).
Larger boxes are in principle possible, but limited by computer time.

\subsection{Global Structure, Flow Collimation, and Jet Formation}

\begin{figure*}[tb]
\plottwo{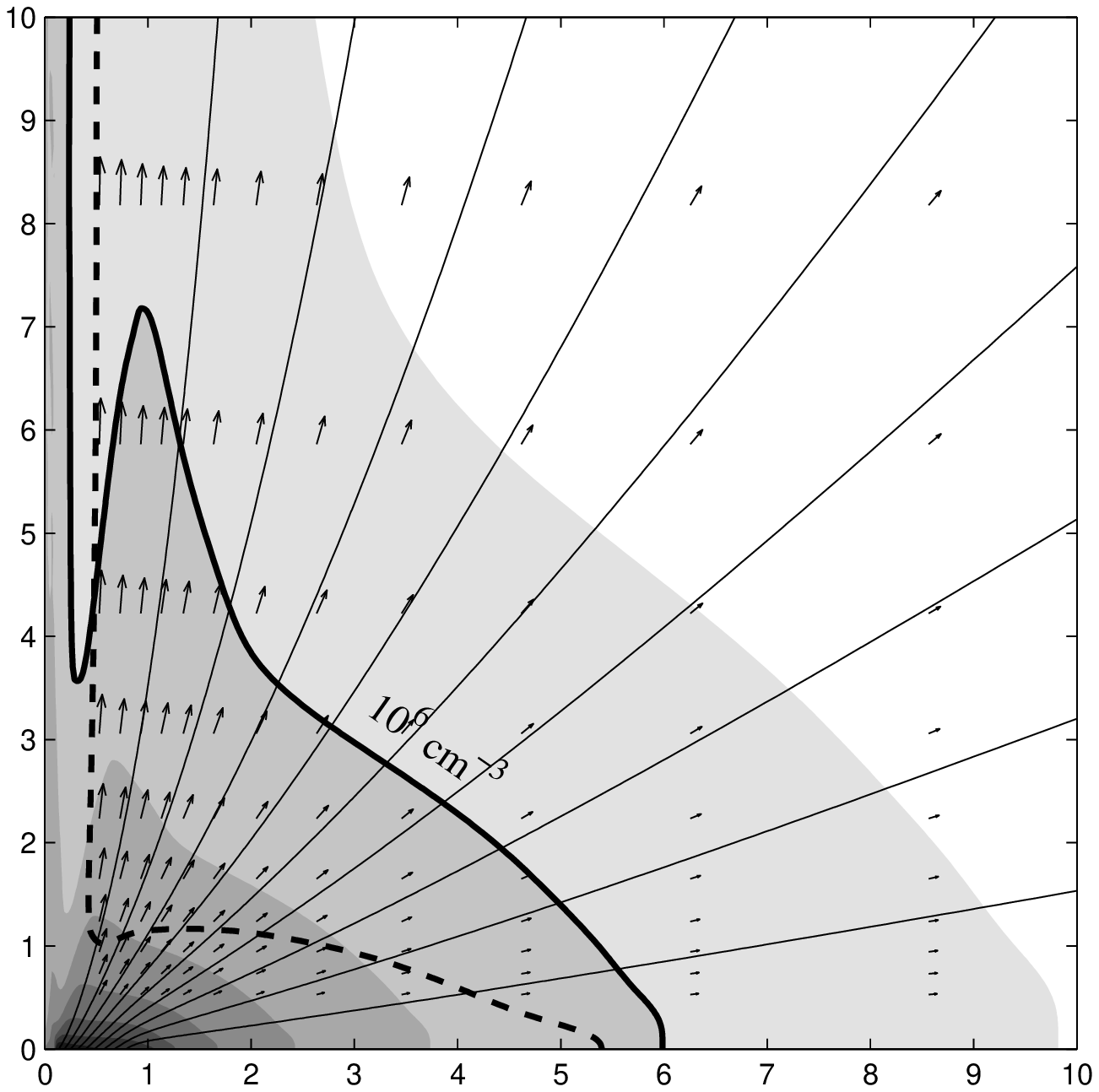}{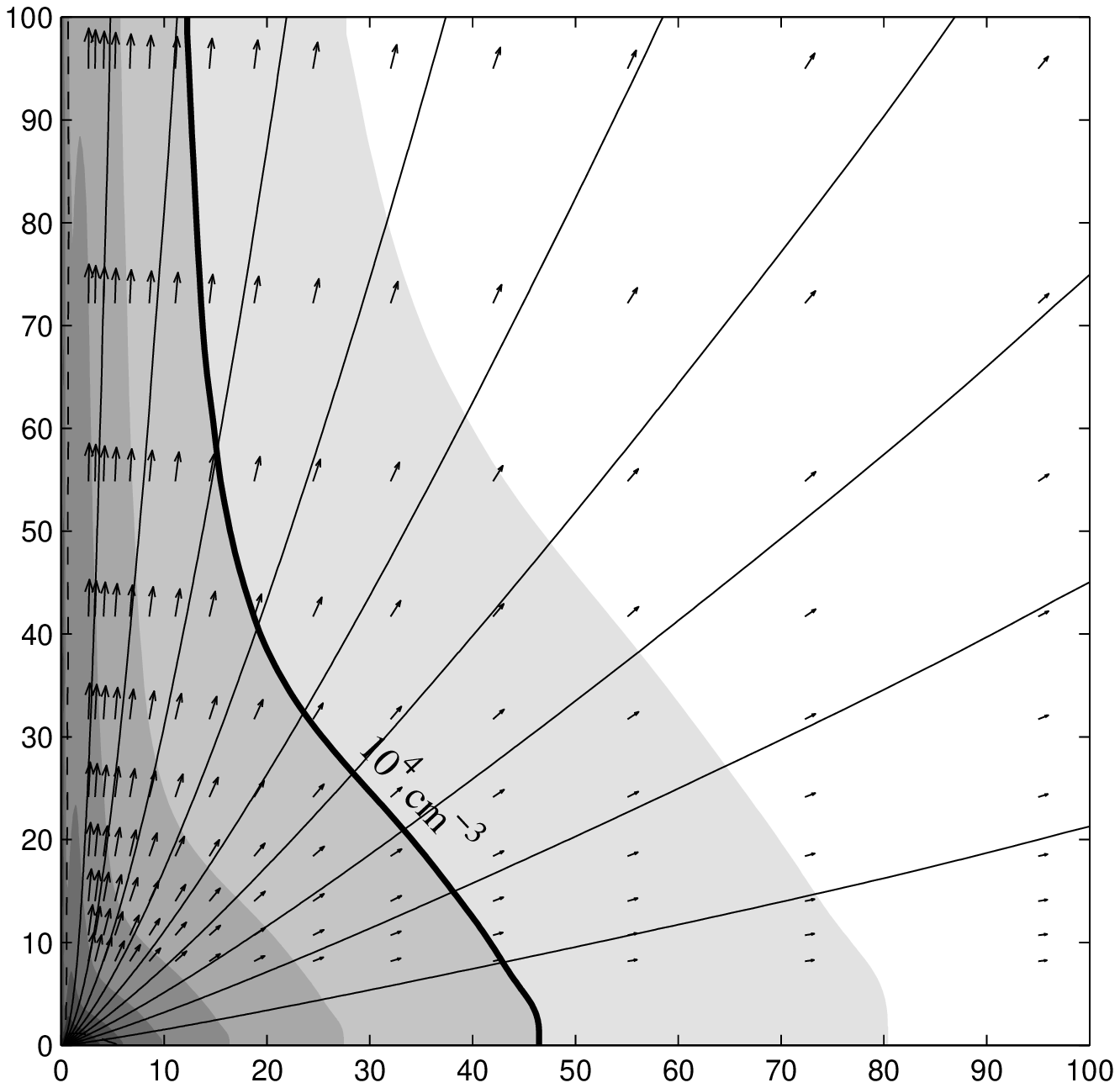}
\caption{Streamlines (light solid) and isodensity contours (heavy solid
lines and shades) of the standard steady wind solution on the (a) $10\AU$
and (b) $10^2\AU$ scale. The streamlines divide the wind into ten zones
of equal mass flux. The dashed line is the fast magnetosonic surface.
The arrows are for poloidal velocity vectors, with length proportional
to the speed. There are two shades per decade in density, and the
densities labeled are the number densities of hydrogen nucleus for a
wind of $10^{-8}\solarmassyr$ (per hemisphere), assuming a helium
abundance of $10\%$ in number.}
\label{fig:4}
\end{figure*}

We follow the evolution of the material injected into the computation
domain numerically with our modified {\software{Zeus}} code,
keeping the boundary
conditions shown in Fig.~\ref{fig:3} fixed at all times. The injected
wind material pushes
aside the low-density ambient medium and propagates towards the outer
edges of the simulation box. After all of the ambient medium is swept
out of the box, the wind settles quickly into a steady state, with
no appreciable time variations. The overall appearance of the
steady-state wind is given in Fig.~\ref{fig:4}, on two
scales. In Fig.~\ref{fig:4}a, we show on the $10\AU$ scale a meridian
view of nine streamlines (also field lines) that divide the wind into
ten zones of equal mass flux. Superposed on the streamlines are the
isodensity contours (in shades), with values of density decreasing
exponentially outwards. A salient feature is the fast magnetosonic
surface, shown in dashed lines in Fig.~\ref{fig:4}a. As mentioned earlier,
the fast surface closes inside the simulation box, except near the axis
where the light, fast axial flow resides. The tenuous flow in this narrow
axial region remains dominated by the (largely poloidal) magnetic field,
which can, in principle, provide some stability to the wind as a whole
against kink instabilities (\citealt{Shu95}). The axial injection
occupies a decreasing fraction of the total volume on larger
scales. It becomes nearly invisible in panel (b) of Fig.~\ref{fig:4},
where streamlines and isodensity contours are plotted on the $100\AU$
scale. Here, most of the space is filled with the flow driven from the
disk magneto-centrifugally, whose acceleration and collimation properties
we want to study in detail.

We first concentrate on the collimation properties of the standard
solution. Self-collimation is commonly accepted as a hallmark of the
magneto-centrifugal winds. The degree of collimation is still a matter
of debate, however (see, e.g., \citealt{Okamoto99}). Our self-consistently
determined solution up to large distances can shed light on this
important issue.
{}From panel (a) of Fig.~\ref{fig:4}, we find that the standard wind is
not well collimated on the small scales close to the launching surface,
although hints of {\it gradual} collimation are evident in the
shapes of both streamlines and density
contours: the streamlines bend slightly towards the rotation axis,
which forces the density contours to elongate along the axis. The
elongation of density contours shows up more clearly on the larger
scale in panel (b) of Fig.~\ref{fig:4}, with the contours appearing
nearly cylindrical in the region within $\sim 15\degree$ of the rotation
axis. Asymptotically cylindrical density stratification was predicted
by \citet{Shu95} analytically. The cylindrical shape does not
extend all the way to the equator, however, at least on scales up
to $100\AU$. The density contours in the equatorial part of the wind
bulge outwards. The bulging is similar to that found numerically by
\citeauthor{Sakurai87}\ (\citeyear{Sakurai87}; see also \citealt{Ouyed97}).
The relative shape
of the bulge appears to change little from one isodensity contour
to the next, which is probably related to the logarithmically slow
collimation of streamlines in the equatorial region at large distances
(see \citealt{Bogovalov99} for a recent discussion). It will be
slow to disappear, if ever, on even larger scales.

The shape of isodensity contours on large scales is important to compute
because it is intimately related to the observed shape of optical jets.
A density stratification bulging out prominently near the base as in
Fig.~\ref{fig:4} may be in conflict with high resolution observations
of some HH jets. In particular, the spectacular HST images of HH 30
(\citealt{Burrows96}; \citealt{Ray96}) show that this jet remains
nearly cylindrical all the way to the disk surface; if anything, it
appears to be slightly pinched in the equatorial region, keeping in
mind, however, that contamination of the jet emission from scattered
stellar light is a concern near the disk. Since the emissivity is
sensitive to
density, the optical appearance of a jet should follow to a large
extent that of isodensity contours \citep{Shang98},
although the thermal structure affecting the emission is
uncertain. The requirement that the density contours be nearly
cylindrical all the way to the disk surface (assuming the likely
situation of a much less rapid variation in temperature than in
density) imposes a constraint on possible
combinations of magnetic field and mass flux distributions on the
launching surface. The standard run does not appear to satisfy this
constraint.

Another generic feature predicted by the asymptotic analysis of \citet{Shu95}
is the strong stratification in density transverse to the rotation
axis. This feature shows up clearly in the density contours. We stress
that the stratification is a most important characteristics of the
magneto-centrifugal wind, even though individual streamlines appear hardly
bent at all. We estimate that, at a height of $90\AU$, the density
distribution decreases rapidly outside a small core of $\sim 2\AU$,
roughly as a power-law of the cylindrical radius $\sim R^{-1.3}$. The
decline is slower than the asymptotic estimate of $\sim R^{-2}$ of
\citet{Shu95}, obtained assuming a constant flow speed on all
field lines; the difference is partly accounted for by the lower speed
achieved along the field lines in the more equatorial part of the
standard solution \citep{Matzner99}.
Nevertheless, the fact that the density decline is significantly faster
than $R^{-1}$ means that the central high-density ``jet'' will stand out
in maps of integrated emission along the line of sight. The steep
density stratification, coupled with a moderate anisotropy in flow velocity,
gives rise to a strong concentration of energy and momentum fluxes in the
direction along the axis, with implications on the dynamics of bipolar
molecular outflows \citep{Li96}, which are probably formed when the
highly anisotropic winds run into the ambient medium. The decrease in
density away from the
axis is qualitatively consistent with the HST observations of
the wind of DG Tau by \citet{Bacciotti00}.

Even though a dense, nearly cylindrical ``jet'' is naturally produced
around the axis, at least well above the disk, the fraction of the
wind mass flux that gets collimated into the jet is a concern. To
fix ideas, we adopt a fiducial
number density of $\nH=10^4\cm^{-3}$ to delineate the outer
boundary of the central ``jet''. From Fig.~\ref{fig:4}b, we find that
only $\sim 20\%$ of the wind mass flux resides in the fiducial jet at
the height of $z=10^2\AU$; the majority remains in the wide-angle
component. Such a jet formation efficiency may be too low, at least for
the famous HH 30 system, as discussed in the introduction. This example
illustrates that there is no guarantee that a disk-driven
magneto-centrifugal wind would automatically (a) concentrate most of
its mass flux into an axial jet or (b) produce a density stratification
cylindrical all the way to the equator.

The relatively low efficiency of jet formation and undesirable shape
of the density contours of the standard solution are related to the
particular distribution of mass flux chosen. It is dominated by the
mass injected at the outer part of the launching region, since the
mass flux per octave in radius is
\begin{equation}
\frac{d {\dot M}}{d \ln R} \propto R^{1/2},
\label{massload}
\end{equation}
between the inner radius of the Keplerian disk $\RI$ and the radius $R_T$
(beyond which the injection speed is modified by equation~[\ref{spline}]).
Mass from the outer
part is loaded onto more horizontally inclined field lines, which are
harder to collimate. The above deficiencies are rectified to some extent
in the wind solutions to be discussed in the next section. Moderately
collimated winds like the standard solution may be applicable to
the outflows from young high-mass stars, which could be more dominated
by the wide-angle component than their low-mass counterparts
\citep{Richer00}.

We note that the standard solution (and other solutions to be discussed
below) does not recollimate towards the rotation axis at large distances,
and does not show any oscillations along the axis. Both behaviors are
seen in some self-similar (\citealt{Chan80}; \citealt{Blandford82};
\citealt{Achtenberg83})
and width-averaged jet solutions (e.g., \citealt{Spruit97}). They
are most likely a result of the self-similarity assumption or the
averaging over the jet cross-section, not the general properties of
magneto-centrifugal winds.

\subsection{Flow Acceleration and Kinetic-Energy Domination}

\begin{figure*}[tb]
\plottwo{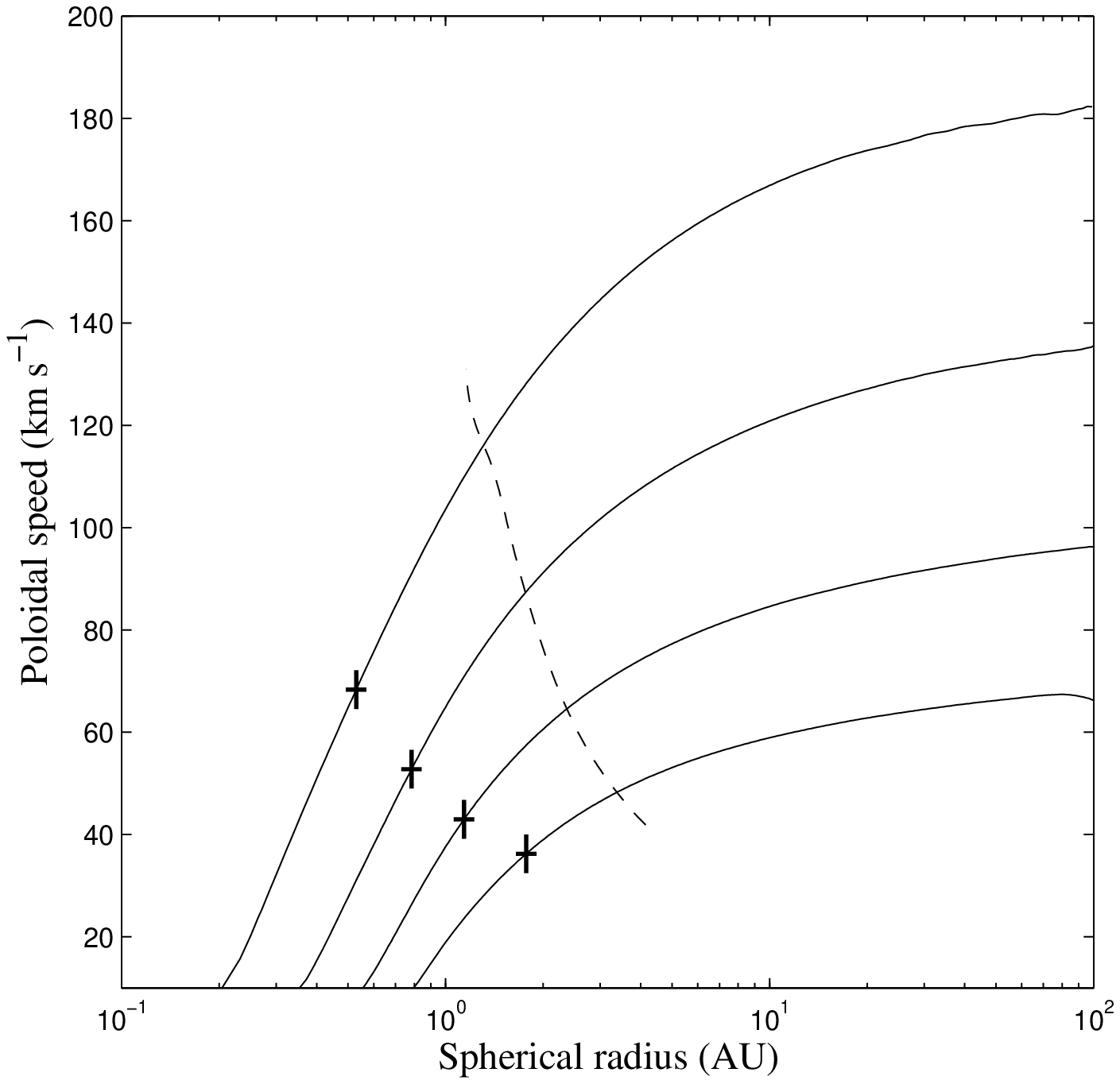}{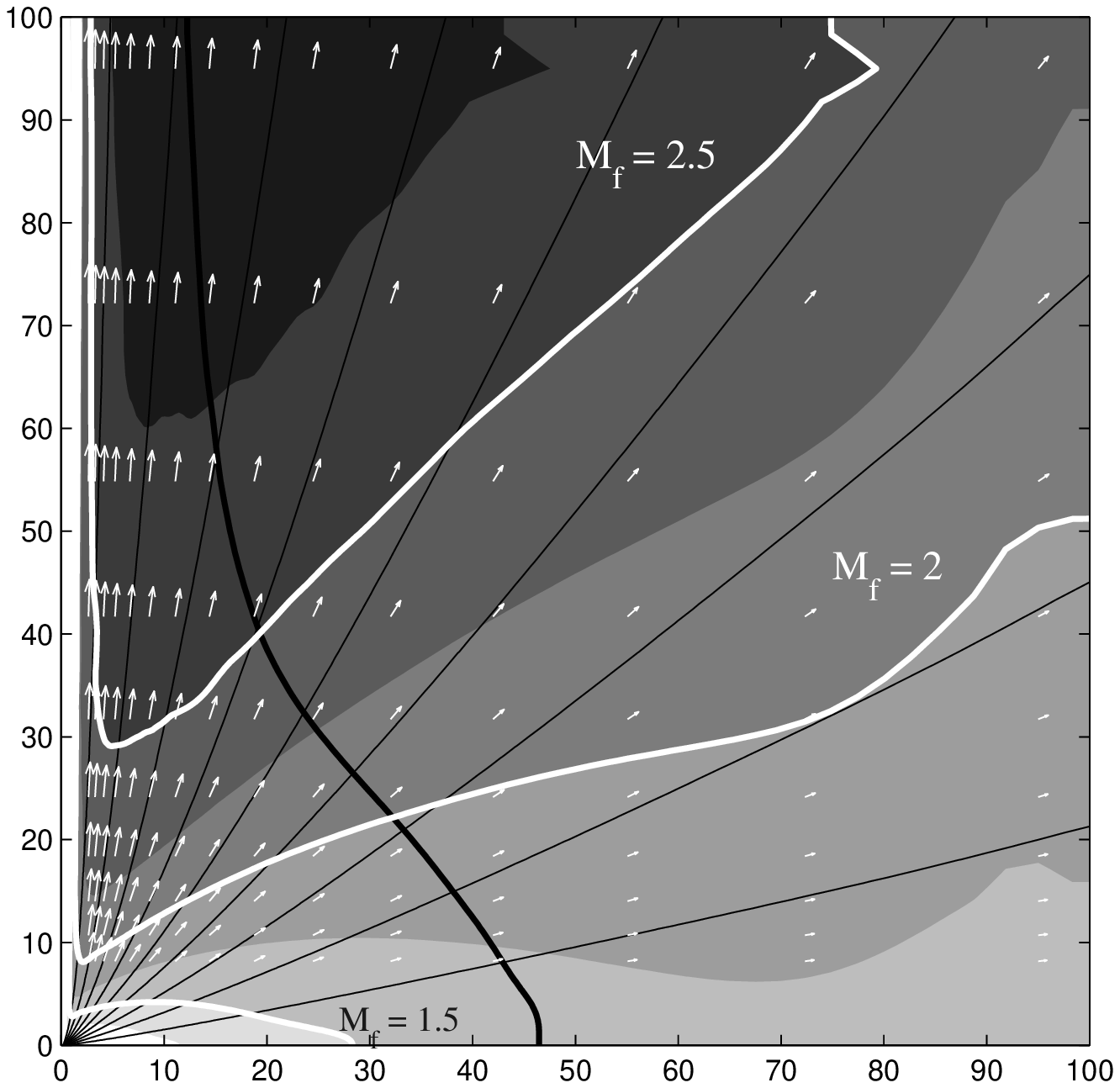}
\caption{Wind acceleration from small to large distances. Plotted are (a)
the poloidal speeds as a function of spherical radius along four field
lines that divide the wind into five zones of equal mass flux, and (b)
the contours of constant fast Mach number $\Mf$,
showing that the wind becomes kinetically dominated at large distances.
Also shown in panel (a) are the fast surface (dashed line) and the
location (cross) along each field line where the poloidal speed matches
the Keplerian speed at the foot point.}
\label{fig:5}
\end{figure*}

We now concentrate on the acceleration, rather than collimation, of the
standard solution. Magneto-centrifugal acceleration along field lines
is illustrated in panel (a) of Fig.~\ref{fig:5}, where the poloidal
velocity is plotted as a function of (log) spherical radius along the
four streamlines that divide the wind into five zones of equal mass
flux. As expected, the flow along each field line crosses the fast point
smoothly, getting accelerated to a speed $\sim 2-3$ times the Keplerian
speed at the foot point of the field line. Note that for this particular
solution, the flow moves faster along the streamlines in the polar region
than in the equatorial region, by a maximum factor of $\sim 4$. This
pole-equator velocity contrast is qualitatively similar to that deduced
for the wind of DG Tau from HST observations \citep{Bacciotti00}.
These authors
inferred that the high velocity component of forbidden line emission
of DG Tau wind occurs near the axis, and is spatially bracketed by
the lower velocity component. Whether wind solutions like our standard
run can explain the density distribution and velocity field observed
quantitatively remains to be seen.

At large distances from the launching region, most of the energy extracted
magnetically from the disk, as measured by the Poynting flux, is expected
to be converted into the kinetic energy of the ordered plasma flow. This
kinetic domination is indeed seen in the standard solution, and is illustrated
in Fig.~\ref{fig:5}b, where contours of constant fast
(magnetosonic Mach) number, $\Mf=v_p/\vf$ (where $v_p$
and $\vf=[(B_p^2+B_\phi^2)/(4\pi\rho)]^{1/2}$ are the poloidal
and fast magnetosonic speed) are plotted. Note that the fast number
along each field line reaches a maximum value between $\sim 1.6$ and
$\sim 3.0$ at the $10^2\AU$ distance from the
origin. Since the ratio of Poynting and kinetic energy fluxes is
approximately $2/\Mf^2$ at large distances \citep{Spruit96},
some $\sim 60\text{--}80\%$ of the total energy is kinetic. Similarly high
(low) kinetic (magnetic) energy fraction has been found numerically
by \citet{Ouyed97} and \citet{Ustyugova99}. This efficient
conversion is of interest because the remaining, low magnetic energy
makes potential destruction of toroidal magnetic field by kink
instabilities (\citealt{Spruit96}; \citealt{Begelman98})
less disruptive to the wind;
the wind has become basically ballistic by the time the outer
edges of the simulation box are reached, particularly in the region
not far from the
polar axis where the fast number is the highest. The transition of the
flow from magnetic to kinetic domination is an esthetically pleasing feature
of the magneto-centrifugal wind. The asymptotic kinetic domination is in
contrast with that of the ``classical'' 1-D Weber-Davis (\citeyear{Weber67})
wind  solution, where the kinetic energy flux is at most a third of the total
\citep{Spruit96}. It also differs from the ultra-relativistic case where
kinetic domination is difficult, if not impossible, to achieve in an
ideal, steady MHD wind that fills the entire $4\pi$ steradians
(\citealt{Chiueh98}; \citealt{Okamoto02}).

We should emphasize that even though the fast number $\Mf$ is
greater than unity, it is not greater by much; it has values of only
a few on the $100\AU$ scale, as seen from Fig.~\ref{fig:5}b. In
other words, the magnetic field, even though its energy fraction is
small, is not completely negligible on large, observable scales. In
particular, the relatively modest values of $\Mf$ make the effects
of magnetic cushion important in the interaction of the magnetized wind
with its ambient medium (or with itself if the wind is variable) through
shocks, especially if the shocks are oblique.

\section{Jets with Better Collimation and Higher Formation Efficiency}
\label{load}

In the standard wind solution discussed in the last section, only a
relatively small fraction of the wind mass flux resides in the
fiducial ``jet'', and the isodensity contours bulge out in the
equatorial region.
Both features are undesirable for modeling jets like HH 30.
In this section, we seek to rectify the situation, by concentrating more
mass loading at smaller disk radii, through a steeper power-law
distribution for the density at the launching surface while keeping
the magnetic field distribution fixed. The density distribution is
controlled by the exponent $e_\rho$ in the equation (\ref{deninj}).
The standard run has a relatively flat density distribution specified
by $e_\rho=1$. We have varied this exponent over a range of values,
and found that the collimation properties of the wind can change
substantially from the standard run. We choose a case with $e_\rho=3$
to illustrate the changes.

\begin{figure*}[tb]
\plottwo{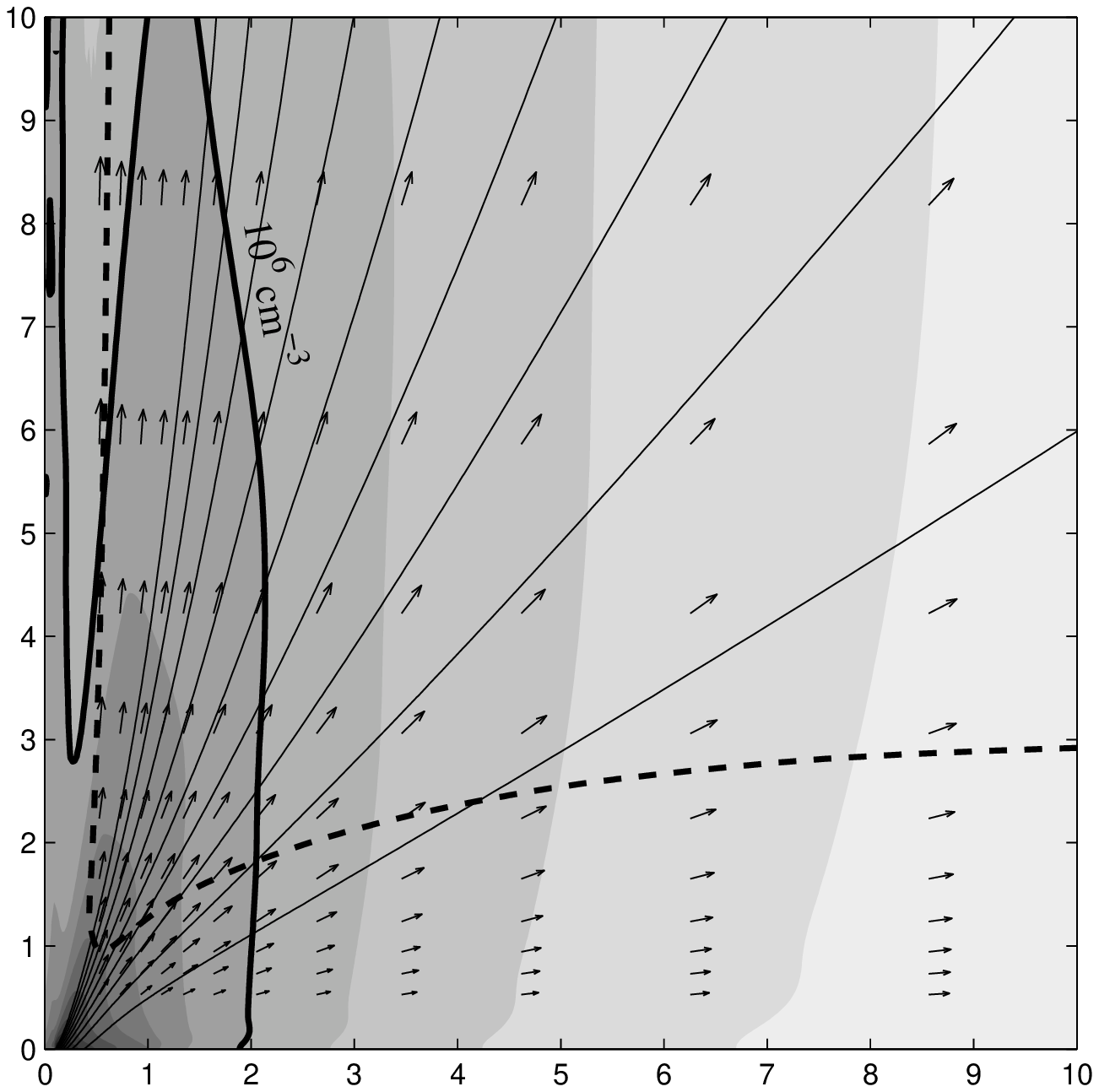}{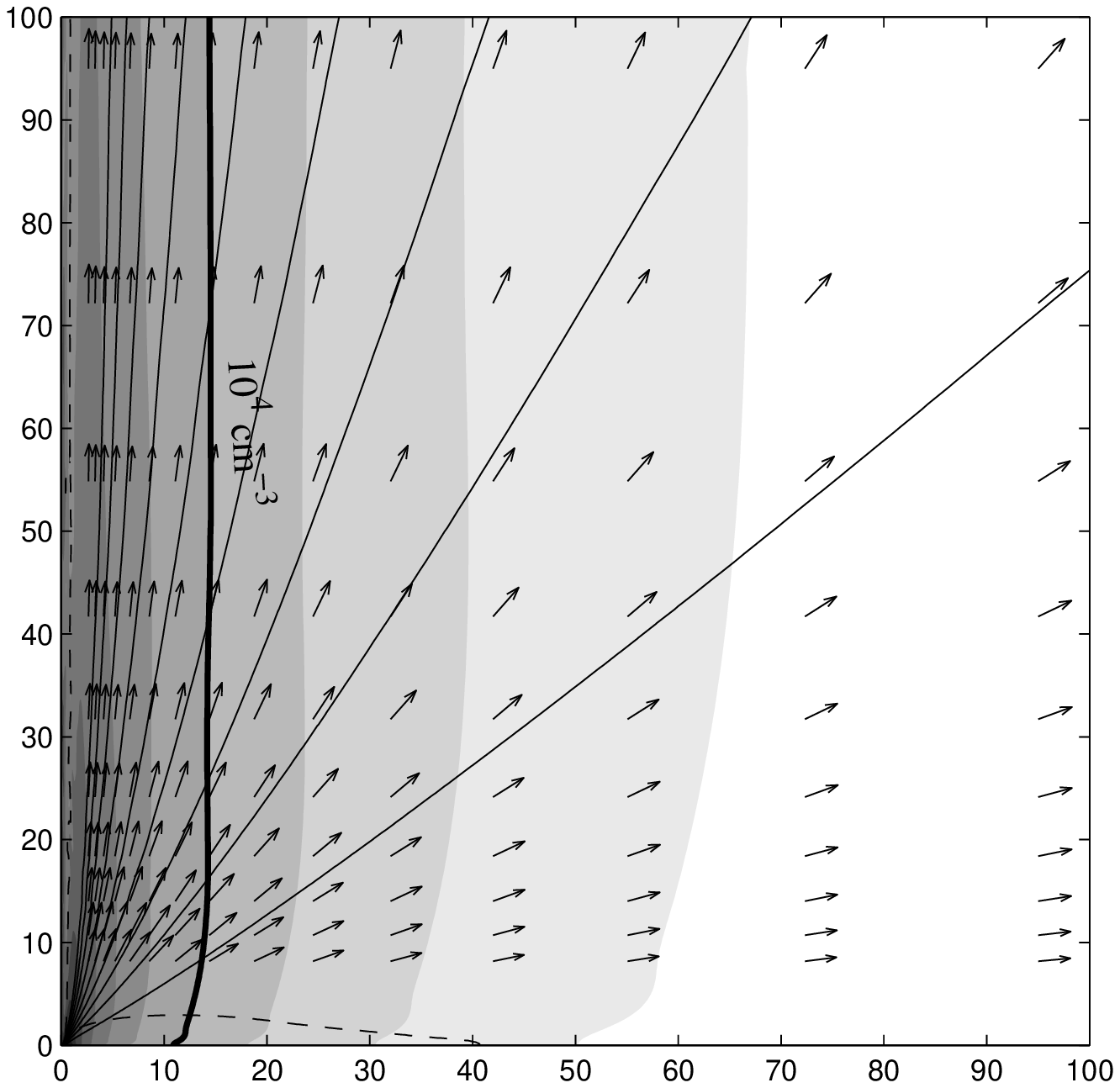}
\caption{Streamlines (light solid) and isodensity contours (heavy solid
lines and shades) of a steady wind solution on the (a) $10\AU$ and
(b) $10^2\AU$ scale, with a mass loading more concentrated near the
inner edge of the Keplerian disk than the standard case. The streamlines
divide the wind into ten zones
of equal mass flux. The dashed line is the fast magnetosonic surface.
The arrows are for poloidal velocity vectors, with length proportional
to the speed. There are two shades per decade in density, and the
densities labeled are the number densities of hydrogen nucleus for a
wind of $10^{-8}\solarmassyr$ (per hemisphere).}
\label{fig:6}
\end{figure*}

For the $e_\rho=3$ run, we keep the distributions of the magnetic
field and injection speed on the disk the same as in the standard
run, and set the density at the inner edge of the Keplerian disk
$\rhoout=10\,\rho_0$ as before. Since the density now
declines more rapidly with radius, the (dimensionless) mass flux
ejected from the Keplerian disk is reduced. We correspondingly reduce
the density
$\rhoin$ at the base of the fast injection region by a factor
of 4 (to $\rhoin=0.025\,\rho_0$), so that its mass flux fraction
remains small ($\sim 2\%$ of the total). The vast majority of the
wind material is launched magneto-centrifugally from the surface
of the Keplerian disk, according to the distribution
\begin{equation}
\frac{d {\dot M}}{d \ln R} \propto R^{-3/2}.
\label{ml2}
\end{equation}
Nearly half of the mass flux is concentrated at the inner edge of the
Keplerian disk, between $1\,\RI$ and $1.5\,\RI$ (or $0.10$ and
$0.15\AU$).
The concentration is shown clearly in panel (a) of Fig.~\ref{fig:6},
where nine streamlines (also field lines) dividing the wind into ten
zones of equal mass flux are plotted, after a steady state is reached.
Most of the streamlines emanate from close to the inner Keplerian
disk, creating an impression of an ``X-wind'', although most of the
wind-launching magnetic flux lies outside the region where the mass
flux is concentrated (see Fig.~\ref{fig:3}b).

The density contours in the $e_\rho=3$ case started out more collimated
in the launching region than those in the standard run. On the
$10^2\AU$ scale
shown in panel (b) of Fig.~\ref{fig:6}, they become nearly cylindrical
all the way to the equator; if anything, the contours pinch slightly
inwards in the equatorial region, in contrast with the bulging observed
in the standard solution. Thus, a better collimated jet is produced.

The efficiency of jet formation is also improved. From Fig.~\ref{fig:6}b,
we estimate that $\sim 45\%$ of the wind mass flux is enclosed, at the
height of $100\AU$, within the density contour of $10^4\cm^{-3}$,
which marks the boundary of the fiducial jet. In contrast, the jet fraction
is only $\sim 20\%$ for the standard run. Therefore, compared with the
standard solution, the wind is more dominated by the axial jet, although
there is still more mass flux in the wide-angle component outside the jet
than inside the jet itself. For the chosen size of the Keplerian disk (one
decade in radius, from $0.1$ to $1.0\AU$) and
magnetic field distribution ($B_z$ roughly proportional to $R^{-3/2}$),
we find it difficult to collimate more than half of the mass flux into
the jet, at least on the $10^2\AU$ scale. The jet fraction may be
increased by external collimation, which can act in addition to the
intrinsic self-collimation. This will be particularly true during the
earliest, Class 0 stage of star formation, when a massive envelope
\citep{Andre00} is present, which can confine the wind. The
combination of external and intrinsic collimation may produce a more
jet-dominated magneto-centrifugal wind; such a wind may be needed
to explain the kinematics of some bipolar molecular outflows \citep{Lee00}.
We will address the issue of external collimation in the future.

We note that the cylindrically-stratified density decreases rapidly
away from the rotational axis, roughly as $R^{-2.2}$ at a height of $90\AU$
in this example.
The decline is steeper than in the standard case, where $\rho\propto
R^{-1.3}$ roughly. It is not far from $R^{-2}$, perhaps signaling that
the wind has reached an asymptotic state close to that analyzed by
\citet{Shu95}. We note that the asymptotic wind speed in this
particular example is almost independent of the polar angle, showing
little signature of the Keplerian rotation at the launching surface. It
illustrates the obvious point
that a disk-wind does not have to be slower in the equatorial
region; it depends on the distribution of mass loading for a given
magnetic field configuration.

As a final remark, we stress that the mass loading in the axial region
where the magneto-centrifugal mechanism fails is somewhat artificial.
Nevertheless, it provides a relatively non-intrusive inner boundary to
the magneto-centrifugally driven outflow. It has the added advantage
of possibly representing a fast stellar coronal wind emanating along
open stellar field lines; such a wind may have a predominantly poloidal
magnetic field, which could add to the stability of the entire flow
\citep{Shu95}.
We have done several tests to ensure the non-intrusive nature of
the fast injection, including increasing by a factor of 4 the density in
the injection region for the two examples discussed in this and last
sections. The flow structure in the magneto-centrifugal region outside
the axial injection remains little affected, especially at large distances.
Even though the axial injection appears to have minimal effects on the
region of interest, it controls the timestep of the simulation through
the Courant condition, because of its low density and large magnetic
field strength (and thus high Alfv{\'e}n speed). To explore parameter
space in a reasonable amount of time, as we do in a companion paper
(J. Anderson et al.\ 2003, in preparation), one can in principle adopt
a relatively heavy axial injection.

\section{Summary and Future Work}
\label{conclusion}

We have developed a {\software{Zeus}}-based numerical code capable of
following the
acceleration and collimation of magneto-centrifugal winds from their
launching surface to large, observable distances. The code is made
possible by treating the wind-driving accretion disk as a boundary,
which allows for a clear separation of the dynamics of the disk and
the wind. By limiting the launching region to the inner part of the
disk, we were able to follow the outflow up to, and well beyond, the
fast magnetosonic surface for most of the streamlines. Close to the
rotation axis where the magneto-centrifugal mechanism fails, a light,
fast jet is injected, which could stay sub-fast magnetosonic in the
computational domain; it occupies an increasingly small fraction of
space at large distances, and provides a non-intrusive inner boundary
to the super-fast magneto-centrifugal wind that we are interested in.

We find that, as expected, most of the magnetic energy extracted from
the disk can be converted into the bulk flow kinetic energy at large
distances, producing an essentially ballistic wind. Its fast magnetosonic
Mach number remains moderate, however, which has implications for wind
interaction. Our unique simulation setup allows us to investigate
quantitatively the large-scale wind structure, particularly jet formation.
In agreement with the asymptotic analysis of \citet{Shu95}, we find
nearly cylindrical stratification in the wind density close to the axis
at relatively large heights. Closer to the equator, the density contours
can either bulge outwards or pinch inwards, depending on the launching
conditions. The fraction of the wind mass flux residing in the dense
axial jet can vary substantially, again depending on the conditions on
the disk, particularly the distribution of mass loading rate for a
given magnetic field configuration. It appears possible to have, at
least on the $10^2\AU$ scales around T Tauri stars, both winds dominated
by the wide-angle component and more jet-dominated winds. The exact
demarcation of these two regimes on the $10^2\AU$ and larger scales is
unclear. It will require a time-consuming, systematic exploration of
parameter space.

Our robust determination of the large-scale structure of magneto-centrifugal
jets and winds opens up several new possibilities.
First, it avoids the artificial instabilities seen in some of previous
simulations where a major fraction of the wind leaves the simulation
box with a sub-fast speed. However, non-axisymmetric instabilities of
physical origin, such as the kink instability, are possible.
They will be examined in a subsequent paper in this series with 3D
calculations (R. Krasnopolsky et al.\ 2003; Paper~III) using
axisymmetric solutions as a starting point.
Second, the size of our wind launching region is adjustable.
It allows us to study both the nominal ``disk-wind''
(e.g.\ \citealt{Konigl00}) and the ``X-wind'' \citep{Shu00},
which differ mainly in the width of the launching region.
The effects of the size of the launching region and other parameters,
such as the mass loading rate, on the large-scale wind structure will
be explored in a companion paper (J. Anderson et al.\ 2003, in preparation).
Finally, by adopting a time-dependent launching condition, we will be
able to study the formation of internal shocks, which may be used to
interpret the knots commonly observed in YSO jets.

\acknowledgments{We thank J. Anderson for helpful discussion and C. Gammie
for computational facilities used in the production runs. The work is
supported in part by NASA grants NAG 5-7007, 5-9180, 5-12102 and by
NSF grant AST 00-93091.}

\end{document}